\documentclass[fleqn, usenatbib]{mnras}

\usepackage[utf8]{inputenc}
\usepackage[T1]{fontenc}
\usepackage{ae,aecompl}

\usepackage{graphicx} 
\usepackage{amsmath}  
\usepackage{amssymb}  

\newcommand{\be}{\begin{equation}}
\newcommand{\ee}{\end{equation}}
\newcommand{\bdm}{\begin{displaymath}}
\newcommand{\edm}{\end{displaymath}}
\newcommand{\bea}{\begin{eqnarray}}
\newcommand{\eea}{\end{eqnarray}}

\newcommand{\msun}{M_\odot}
\def\lsim{\lower.5ex\hbox{$\; \buildrel < \over \sim \;$}}

\usepackage[usenames,dvipsnames]{xcolor}


\begin{document}

\title[eLISA measurements of eccentric binaries] {\textbf Constraining
  stellar binary black hole formation scenarios with eLISA
  eccentricity measurements}

\author[A. Nishizawa et al.]
       {Atsushi Nishizawa $^{1}$\thanks{E-mail: anishiza@olemiss.edu}, Alberto Sesana, $^{2}$, Emanuele Berti$^{1,3}$, Antoine Klein$^{1,3}$ 
\\
%
$^{1}$ Department of Physics and Astronomy, The University of 
Mississippi, University, MS 38677, USA\\
$^{2}$ School of Physics and Astronomy, University of
Birmingham, Edgbaston, Birmingham B15 2TT, United Kingdom\\
$^{3}$ CENTRA, Departamento de F\'isica, Instituto Superior
T\'ecnico, Universidade de Lisboa, Avenida Rovisco Pais 1,
1049 Lisboa, Portugal}

\date{\today}

\pagerange{\pageref{firstpage}--\pageref{lastpage}} \pubyear{2016}

\maketitle

\label{firstpage}

\begin{abstract}
  A space-based interferometer such as eLISA could observe few to few
  thousands progenitors of black hole binaries (BHBs) similar to those
  recently detected by Advanced LIGO. Gravitational radiation
  circularizes the orbit during inspiral, but some BHBs retain a
  measurable eccentricity at the low frequencies where eLISA is most
  sensitive. The eccentricity of a BHB carries precious information
  about its formation channel: BHBs formed in the field, in globular
  clusters, or close to a massive black hole (MBH) have distinct
  eccentricity distributions in the eLISA band. We generate mock eLISA
  observations, folding in measurement errors, and using Bayesian
  model selection we study whether eLISA measurements can identify the
  BHB formation channel. We find that a handful of observations would
  suffice to tell whether BHBs were formed in the gravitational field
  of a MBH. Conversely, several tens of observations are needed to
  tell apart field formation from globular cluster formation. A
  five-year eLISA mission with the longest possible armlength is
  desirable to shed light on BHB formation scenarios.
 \end{abstract}

\begin{keywords}
black hole physics - gravitational waves
\end{keywords}

\section{Introduction}
The first direct observation of merging black hole binaries (BHBs)
during the first observation run (O1) of Advanced LIGO marked a
milestone in the history of astronomy and fundamental physics.  The
observation of two events (GW150914 and GW151226), plus a third
candidate LVT151012
\citep{2016PhRvL.116f1102A,2016arXiv160604855T,2016arXiv160604856T},
provides a formidable laboratory to test general relativity in the
strong-gravity regime \citep{2016PhRvL.116v1101A}. In addition,
gravitational-wave observations of BHBs can further our
understanding of their astrophysical formation
channels~\citep{2016ApJ...818L..22A}.
Different formation scenarios result in distinct BHB binary
properties, that may be disentangled by analysing the statistical
parameters of a sufficiently large number of detections. The currently
favoured scenarios involve stellar evolution of field binaries
\citep{2013LRR....16....4B} and the dynamical capture of BHs in
globular clusters \citep{2014LRR....17....3P}. Recent work showed that
both field formation
\citep{2009MNRAS.395L..71M,2010ApJ...715L.138B,2012ApJ...759...52D,2013ApJ...779...72D,2015MNRAS.451.4086S,2015ApJ...806..263D,2016arXiv160204531B,2016ApJ...819..108B}
and cluster formation
\citep{2015PhRvL.115e1101R,2016PhRvD..93h4029R,2016arXiv160300884C,2016ApJ...824L...8R}
are compatible with the recent Advanced LIGO
observations~\citep{2016ApJ...818L..22A}. More exotic proposals
include formation via hierarchical triples in the background of a
MBH~\citep{Antonini:2015zsa}, a Population III origin for the binary
members \citep{2014MNRAS.442.2963K,2016MNRAS.tmpL..54H}, chemically
homogeneous evolution in short-period
binaries~\citep{2016MNRAS.458.2634M,2016MNRAS.tmp..892D}, massive
overcontact binary evolution \citep{2016A&A...588A..50M}, and even
primordial BHs
\citep{2016PhRvL.116t1301B}.

In the field formation scenario BHBs have very small eccentricities in
the Advanced LIGO band, and the BH spins should be preferentially
aligned with the orbital angular momentum as a consequence of mass
transfer episodes preceding the stellar collapse. In the globular
cluster scenario BHBs are formed via BH capture on eccentric orbits,
and the spins of the two BHs are likely to be randomly
oriented. Earth-based gravitational-wave observations could
potentially differentiate between field and cluster formation by
looking at spin dynamics, redshift distribution and possibly kicks.
However the details of mass transfer and tidal alignment in BH
binaries, as well as the degree of asymmetry in stellar collapse --
and the resulting kicks imparted to the BHs -- are quite uncertain,
and they will affect BH spin alignment and gravitational waveforms in
complex ways
\citep{2008ApJ...682..474B,Gerosa:2013laa,2015PhRvD..92f4016G,2016arXiv160204531B}.
In this respect, eccentricity may be a more robust tracer of the
formation channel. Radiation reaction is well known to circularize the
orbit. While field and cluster formation scenarios predict very
distinct eccentricity distributions at BHB formation, both scenarios
result in nearly circular binaries in the Advanced LIGO band.  The
first observed signals did not set strong bounds on the eccentricity
of the binary \citep{2016arXiv160203840T,2016arXiv160604856T},
and it is quite unlikely that eccentricity measurements with
ground-based detectors will ever differentiate between the field and
cluster scenarios. However, \cite{2016PhRvL.116w1102S} showed that,
depending on the intrinsic rates (which are only loosely constrained
by current detections) and on the detector baseline, the evolved Laser
Interferometer Space Antenna (eLISA) will observe few to few thousands
BHBs \citep[see also][]{2016arXiv160602298K}.  Because of the much
lower frequency band, eLISA will detect these systems {\it before}
circularization, and in many cases it will be able to measure their
eccentricity \citep{2016arXiv160501341N}.

In this Letter we use Bayesian model selection to demonstrate how
eLISA eccentricity measurement can conclusively distinguish between
different BHB formation channels. In Section II we consider three
models for BHB formation, and discuss the eccentricity distributions
predicted by these models in the eLISA band\footnote{For a detailed
  astrophysical comparison of BHBs formed in galactic fields and
  globular clusters observable by eLISA, see \cite{Breivik}.}.  In
Section III we simulate and analyse eLISA observations using various
models and detector baselines. In Section IV we present our main
results, and in Section V we discuss their implications. We assume a
concordance $\Lambda$CDM cosmology with $h=0.679$, $\Omega_M=0.306$
and $\Omega_\Lambda=0.694$ \citep{2015arXiv150201589P}.

\section{Astrophysical models and eccentricity distributions}
\label{sec2}

We consider a BHB population merging at a rate ${\cal R}$,
characterized by a chirp mass probability distribution $p({\cal M}_r)$
-- where
${\cal M}_r\equiv (M_{1,r}M_{2,r})^{3/5}/(M_{1,r}+M_{2,r})^{1/5}$, and
a subscript $r$ denotes quantities in the rest frame of the source --
and by an eccentricity probability distribution $p(e_*)$ at some
reference frequency $f_*$ close to coalescence (we set $f_*=10$Hz).
If $p(e_*)$ depends only on the BHB formation route, but not on chirp
mass and redshift, the merger rate density per unit mass and
eccentricity is given by
\begin{equation}
  \frac{d^3n}{d{\cal M}_rdt_rde_*}= p({\cal M}_r)\,p(e_*)\,{\cal R}.
  \label{eqrate2}
\end{equation}
Equation (\ref{eqrate2}) can be then converted into a number of sources emitting  per unit mass, redshift and frequency at any time via
\begin{equation}
  \frac{d^4N}{d{\cal M}_rdzdf_rde_*}=\frac{d^3n}{d{\cal M}_rdt_rde_*}\frac{dV}{dz}\frac{dt_r}{df_r}(e(e_*,f)),
  \label{eq1}
\end{equation}
where $dV/dz$ is the standard volume shell per unit redshift, and 
\begin{equation}
  \frac{dt_r}{df_r}(e(e_*,f))=\frac{5c^5}{96\pi^{8/3}}(G{\cal M}_r)^{-5/3}f_r^{-11/3}\frac{1}{F(e(e_*,f))}.
  \label{eq2}
\end{equation}
Here
\begin{equation}
  F(e(e_*,f))=(1-e^2)^{-7/2}\left(1+\frac{73}{24}e^2+\frac{37}{96}e^4\right)\,,
  \label{Fe}
\end{equation}
and $e(e_*,f)$ is computed by finding the root of 
\begin{equation}
  \frac{f}{f_*}=\left[\frac{1-e_*^2}{1-e^2}\left(\frac{e}{e_*}\right)^{12/19}\left(\frac{1+\frac{121}{304}e^2}{1+\frac{121}{304}e_*^2}\right)^{870/2299}\right]^{-3/2}.
  \label{eqe_implicit} 
\end{equation}
We can construct a population of systems potentially observable by
eLISA by Monte Carlo sampling from the distribution in equation
(\ref{eq1}) using appropriate distribution functions for
$p({\cal M}_r)$ and $p(e_*)$. For the mass distribution we employ the
``flat'' mass function of \cite{2016ApJ...818L..22A}, i.e., we assume
that the two BH masses are independently drawn from a log-flat
distribution in the range $5\msun<M_{1,2,r}<100\msun$, restricting the
total BHB mass to the be less than $100\msun$. For the eccentricity
distribution we consider, as a proof of concept, three popular BHB
formation scenarios:
\begin{enumerate}
\item Model {\it field}: this is the default BHB field formation
  scenario of ~\cite{Kowalska:2010qg}, taken to be representative of
  BHBs resulting from stellar evolution.
\item Model {\it cluster}: globular clusters efficiently form BHBs via
  dynamical capture.  Most of these BHBs are ejected in the field and
  evolve in isolation until they eventually merge. Because of their
  dynamical nature, BHBs typically form with a thermal eccentricity
  distribution. A comprehensive study of this scenario has been
  performed by \cite{Rodriguez:2016kxx}.
\item Model {\it MBH}. BHs and BHBs are expected to cluster in
  galactic nuclei because of strong mass segregation. In this case,
  binaries within the sphere of influence of the central MBH undergo
  Kozai-Lidov resonances, forming triplets in which the external
  perturber is the MBH itself. This scenario has been investigated in
  \cite{Antonini:2012ad}, and it results in high BHB eccentricities.
\end{enumerate}

\begin{figure}
\begin{center}
\includegraphics[width=7.5cm]{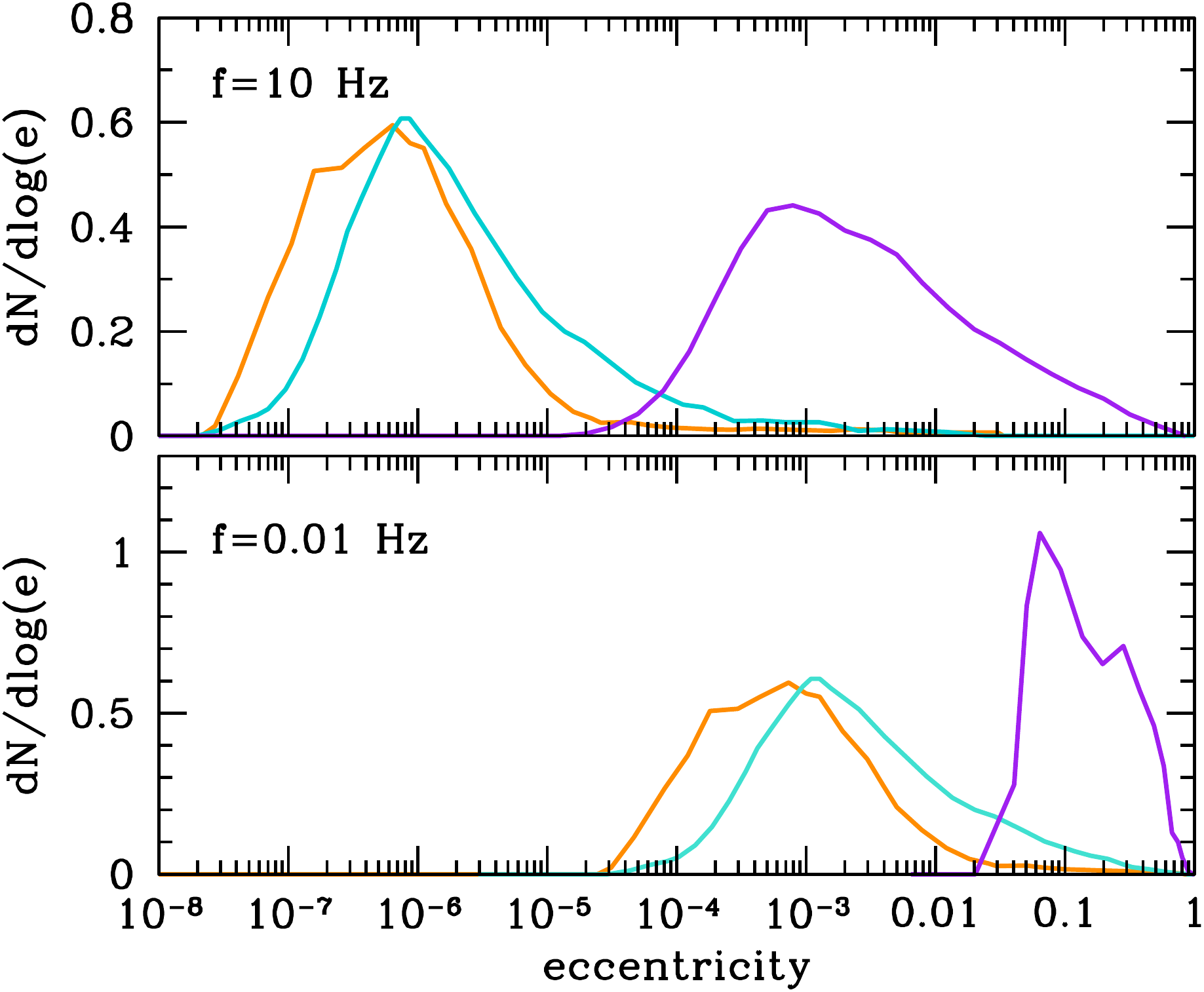}
\caption{Eccentricity distributions predicted by the {\it field}
  (orange), {\it cluster} (turquoise) and {\it MBH} (purple)
  scenarios. The top panel show the distribution at the reference
  frequency $f_*=10$Hz, while the bottom panel is the observable
  distribution $p(e_0)$ evolved ``back in time'' to $f_0=0.01$Hz.}
\label{fig:edist}
\end{center}
\end{figure} 
 
The eccentricity distributions at $f_*=10$Hz, as predicted by these
models, are shown in the top panel of Figure~\ref{fig:edist}. In the
bottom panel we propagate these distributions ``back in time'' to
obtain $p(e_0)$ at frequency $f_0=0.01$Hz, where most eLISA detections
are expected to occur.  In this calculation we must take into account
the fact that highly eccentric binaries evolve more quickly -- by a
factor $F(e)$ -- than circular ones, so that only a few highly
eccentric binaries will be observable in the eLISA band for a given
coalescence rate.

\section{Simulations and analysis tools}
\label{sec3}

We consider two eLISA baselines, N2A2 and N2A5 in the notation of
\cite{2016PhRvD..93b4003K}.  We adopt the noise level (N2) recently
demonstrated by LISA Pathfinder \citep[LPF,][]{2016PhRvL.116w1101A}
and, following the recommendations of the GOAT
committee\footnote{\url{http://www.cosmos.esa.int/web/goat}}, we choose
armlengths of two (A2) and five (A5) million kilometers.  We also
explore two nominal mission lifetimes (2 and 5 years) for a total of
four mission baselines: N2A2-2y, N2A2-5y, N2A5-2y, N2A5-5y.

For our simulated experiments we need (a) the {\it theoretical}
eccentricity distribution predicted by the three BHB formation models,
and (b) catalogues of ``synthetic'' eLISA observations to be tested
against the models. The theoretical distributions are generated as
follows:
(i) Following the formalism described in Section \ref{sec2}, for each
model we generate a large Monte Carlo catalogue of BHBs emitting in
the eLISA frequency window; (ii) For each eLISA baseline, we select a
sample of $10^4$ detectable BHBs with $S/N>8$ and construct their
eccentricity distribution; (iii) We fit a smooth polynomial function
to each distribution.  This procedure yields 12 {\it theoretical}
eccentricity distributions to be tested against observations for each
setup (i.e., for each combination of BHB formation model and eLISA
baseline).

\begin{figure}
\begin{center}
\includegraphics[width=\columnwidth]{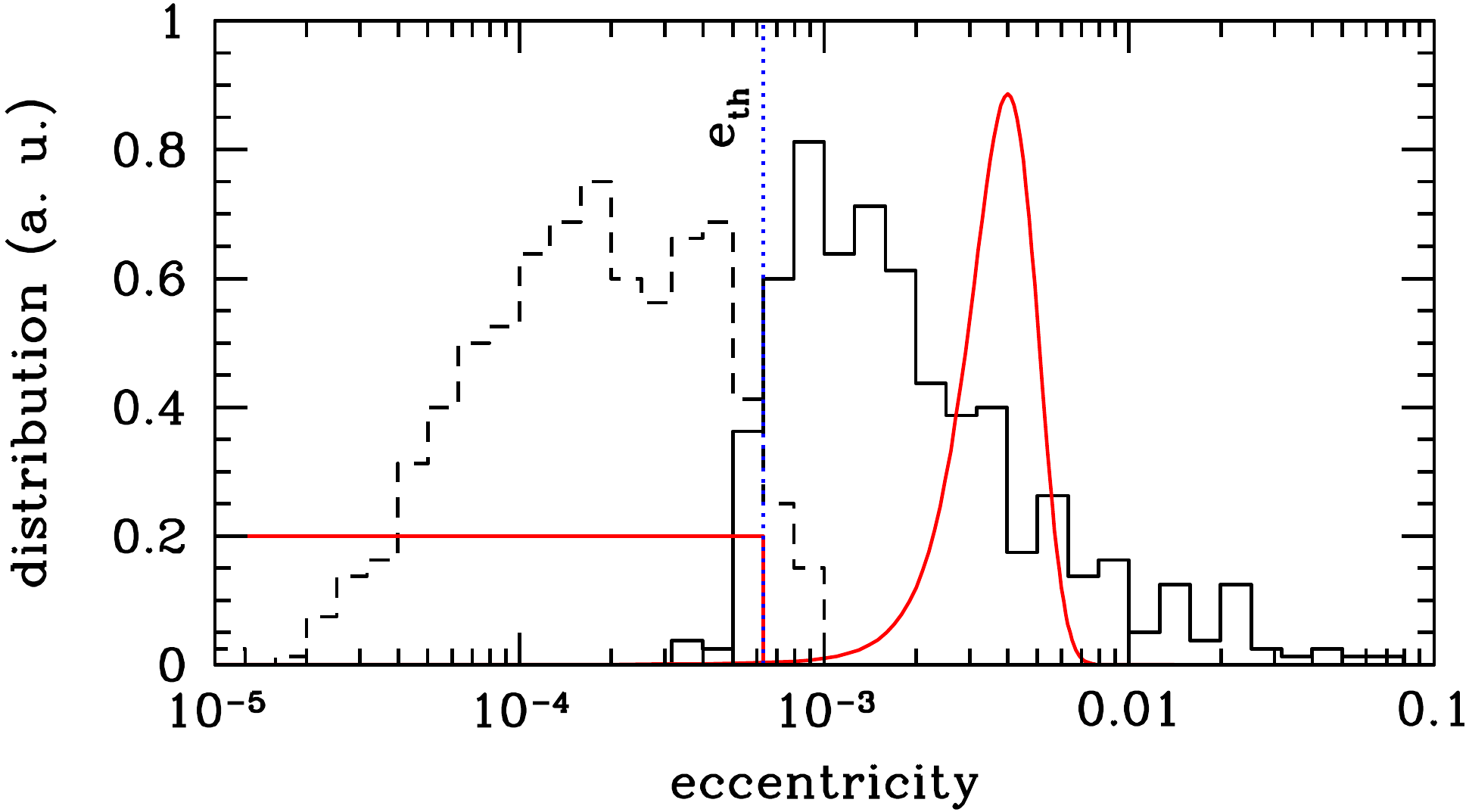}
\caption{Distribution of BHBs with $\Delta e_0<e_0$ (solid histogram)
  and with $\Delta e_0>e_0$ (dashed histogram), assuming the {\it
    field} formation channel and the N2A5-5yr baseline. The dashed
  blue vertical line marks $e_{\rm th}\approx 6\times10^{-4}$. Two
  examples of $p(e_0)$ are shown in red: a Gaussian distribution for a
  case with measured eccentricity (right), and a step function
  for an upper limit (left).}
\label{fig:example}
\end{center}
\end{figure}

The next step is the generation of synthetic BHB observations.  We
select $10^3$ detectable events with $S/N>8$ (because of computational
limitations) and with eccentricity smaller than $0.1$ when eLISA
observations begin (because of limitations in our waveform model, that
becomes unreliable for high eccentricities).  For each of these events
we compute the error $\Delta e_0$ in the measurement of the binary
eccentricity $e_0$ using the Fisher information matrix, as described
in \cite{2016arXiv160501341N}.
For any given number $1\leq N_{\rm obs}\leq 100$ of observed BHBs, we
draw 100 random catalogues of $N_{\rm obs}$ systems from the $10^3$
simulated detections.

As shown by \cite{2016arXiv160501341N}, eLISA will enable precise
measurements of $e_0$ down to $e_0\approx 10^{-3}$, with mild
dependence on armlength or observation time.  Roughly speaking, as
shown in Figure~\ref{fig:example}, this means that $e_0$ can be
measured above some threshold $e_{\rm th}$.
To fold in error measurements into each catalogue, we split the events
in two classes: 1) if $\Delta e_0<e_0$ the eccentricity is measurable,
and we assign to the eccentricity a probability distribution
$p(e_0)=1/\sqrt{2\pi\Delta e_0^2}\times{\rm
  exp}[-(e_0-\bar{e})^2/(2\Delta e_0^2)]$,
where the measured value $\bar{e}$ is drawn from a Gaussian
distribution centred in $e_0$ with variance $\Delta e_0$; 2) if
$\Delta e_0>e_0$ we simply assume that we have an upper limit on
$e_0$, that we take to be $e_{\rm th}$ for simplicity. Therefore
$p(e_0)$ is just a step function: $p(e_0)=1/e_{\rm th}$ for
$e_0<e_{\rm th}$, and $p(e_0)=0$ for $e_0>e_{\rm th}$. Examples of this
procedure are shown in Figure~\ref{fig:example}.

For each setup and each $N_{\rm obs}$ this procedure yields 100
catalogues of observed events, each characterized by the appropriate
$p(e_0)$. The 100 catalogues are not independent when
$N_{\rm obs}>10$, since most of them will share some events. However,
even for the ``shared'' events $\bar{e}$ is obtained from a different
random draw each time, and therefore $p(e_0)$ is different in
different catalogues. We checked that the results presented in the
next section are unaffected when we consider a smaller number of truly
independent catalogues (e.g., 10 catalogues with $N_{\rm obs}=100$).
We are therefore confident that our results are robust with respect to
statistical fluctuations.

\subsection{Statistical analysis}

For a given eLISA baseline, our main goal is to consider catalogues of
$N_{\rm obs}$ observed events from model $A$ (chosen among {\it
  field}, {\it cluster}, {\it MBH}) and assess to what confidence (if
any) we can say whether the underlying astrophysical model was in fact
$A$, or an alternative model $B$.
This is a classic Bayesian model selection problem. To compare two
models $A$ and $B$ we must compute the odds ratio
\begin{equation}
 O_{AB}= \frac{{\cal Z}_AP(A)}{{\cal Z}_BP(B)},
  \label{eqodds}
\end{equation}
where ${\cal Z}_X$ is the evidence of model $X$ and $P(X)$ is the prior
belief that model $X$ is right. In absence of prior information, we
conservatively assume $P(A)=P(B)=0.5$. Moreover, for non-parametric
models (i.e. models without free parameters, as those considered
here), the evidence is simply the likelihood $P(D|X)$ of the data
given the model. The odds ratio then reduces to the likelihood ratio
\begin{equation}
 O_{AB}= \frac{P(D|A)}{P(D|B)}.
  \label{eqodds2}
\end{equation}
In our case, for each set of $N_{\rm obs}$ measurements we have
$i=1,\,\dots,\,N_{\rm obs}$ probability distributions of measured eccentricity
$p_i(e_0)$ to be compared to the theoretical eccentricity distribution
predicted by a given model, ${\cal P}_X(e_0)$. The likelihood of the
data given the model is therefore given by:
\begin{equation}
 P(D|X)=\prod_{i=1}^{N_{\rm obs}}\int p_i(e_0){\cal P}_X(e_0)de_0.
  \label{eqlike}
\end{equation}
In this framework, we can also assign to model $A$ a probability
$p_A=P(D|A)/[P(D|A)+P(D|B)]$ and to model $B$ the complementary
probability $p_B=1-p_A$. From equation (\ref{eqodds2}),
$p_A=O_{AB}/(O_{AB}+1)$. It is therefore natural to associate the odds
ratio with the ``confidence'' in a given model. For example, if
$O_{AB}\simeq 20$ then $p_A\simeq 0.95$, and we can say that model $A$
is favoured at $2\sigma$ (95\% confidence). Definitive $5\sigma$
identification can be associated to $O_{AB}\sim 10^6$.

For each of the 100 catalogues of $N_{\rm obs}$ detections, generated
for each setup, we compute the likelihood of the observed data against
each model: {\it field}, {\it cluster}, {\it MBH}. If the observations
can discriminate between different models, the odds ratio will favour
the actual model from which the data were drawn. For each comparison
we then compute the probabilities $p_A$ and $p_B$ defined above, which
describe our degree of confidence that the data were actually drawn
from either of the models considered in the comparison.

\section{Results}

\begin{figure}
\begin{center}
\includegraphics[width=\columnwidth]{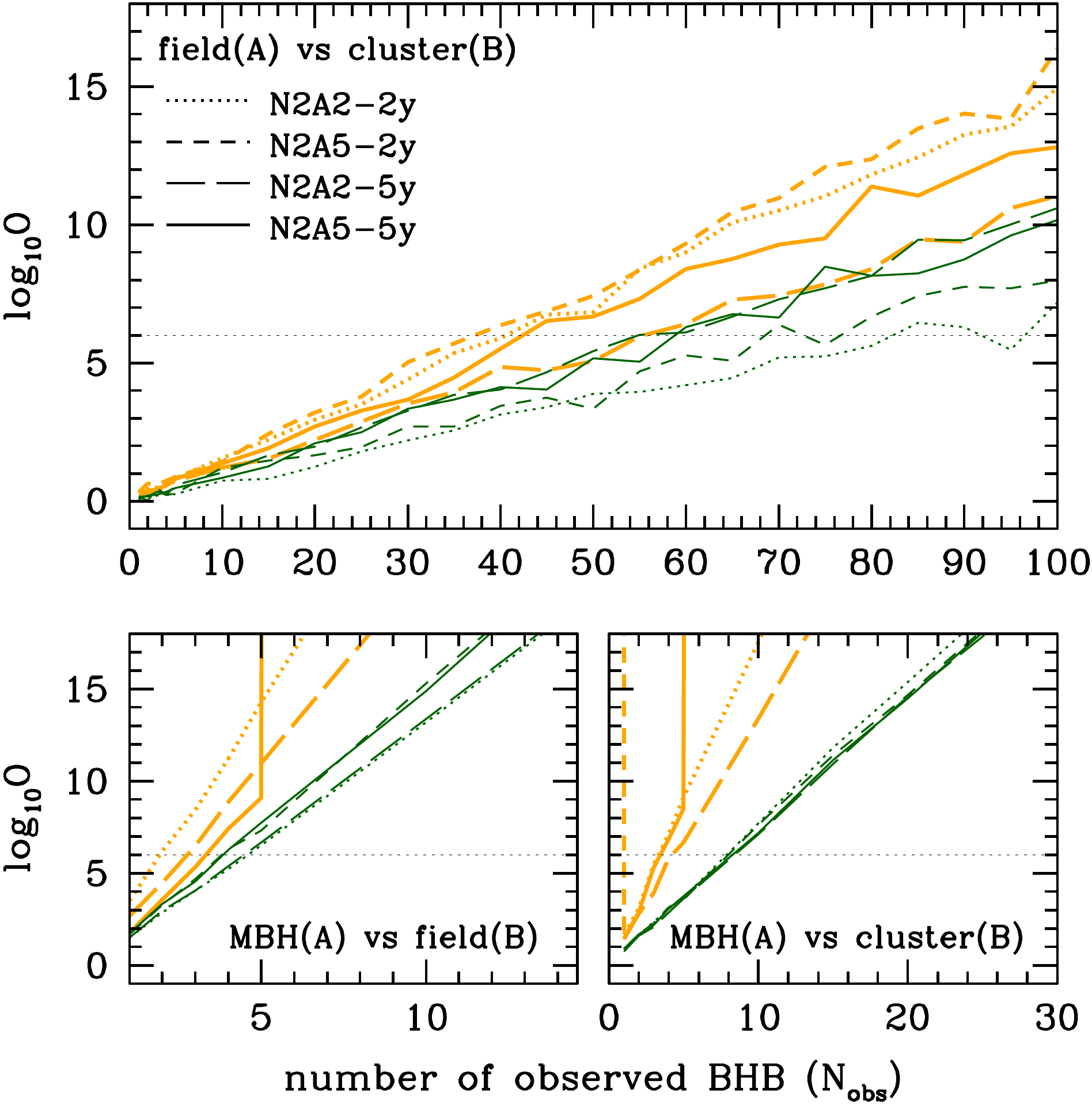}
\caption{Median odds ratio as a function of $N_{\rm obs}$ for
  different model pairs and different detector baselines, as labeled
  in the figure. In each comparison, thick orange curves represent $O_{AB}$
  when $A$ is the true model, whereas thin green curves represent $O_{BA}$
  when $B$ is the true model.}
\label{fig:oddsall}
\end{center}
\end{figure}

In Figure~\ref{fig:oddsall} we compare the median log$(O_{AB})$ as a
function of $N_{\rm obs}$ for each pair of models. The median is
computed over 100 Monte Carlo catalogues for each value of
$N_{\rm obs}$. The two lower panels show that, regardless of the
detector baseline, model {\it MBH} can be confidently separated --
with $\log(O)=6$, i.e. at approximately $5\sigma$ -- from any other
model after a handful of observations. This is because the
eccentricity distribution for model {\it MBH} is biased towards high
values (see Figure~\ref{fig:edist}), at variance with other
models. Note also that it is easier to reject model {\it MBH} when it
is false (orange curves) than to confirm it when it is true (green
curves). This is because models {\it field} and {\it cluster} allow
for low eccentricities, that are not supported by model {\it MBH}
(Figure~\ref{fig:edist}). As soon as a BHB has a measured eccentricity
$e_0<0.01$, model {\it MBH} is automatically rejected. The converse is
not true: the eccentricity range of model {\it MBH} is also supported
by the other models. Therefore, when BHB formation is indeed described
by model {\it MBH} there is always a chance that highly eccentric BHBs
were drawn from other models, and a few more detections are required
to reject them. Models {\it field} and {\it cluster}, on the other
hand, predict similar eccentricity distributions,. Depending on which
one was the true model and on the eLISA baseline, a number of
detections between 30 and 95 is needed to achieve the $\log(O)=6$
threshold.

\begin{figure}
\begin{center}
\includegraphics[width=\columnwidth]{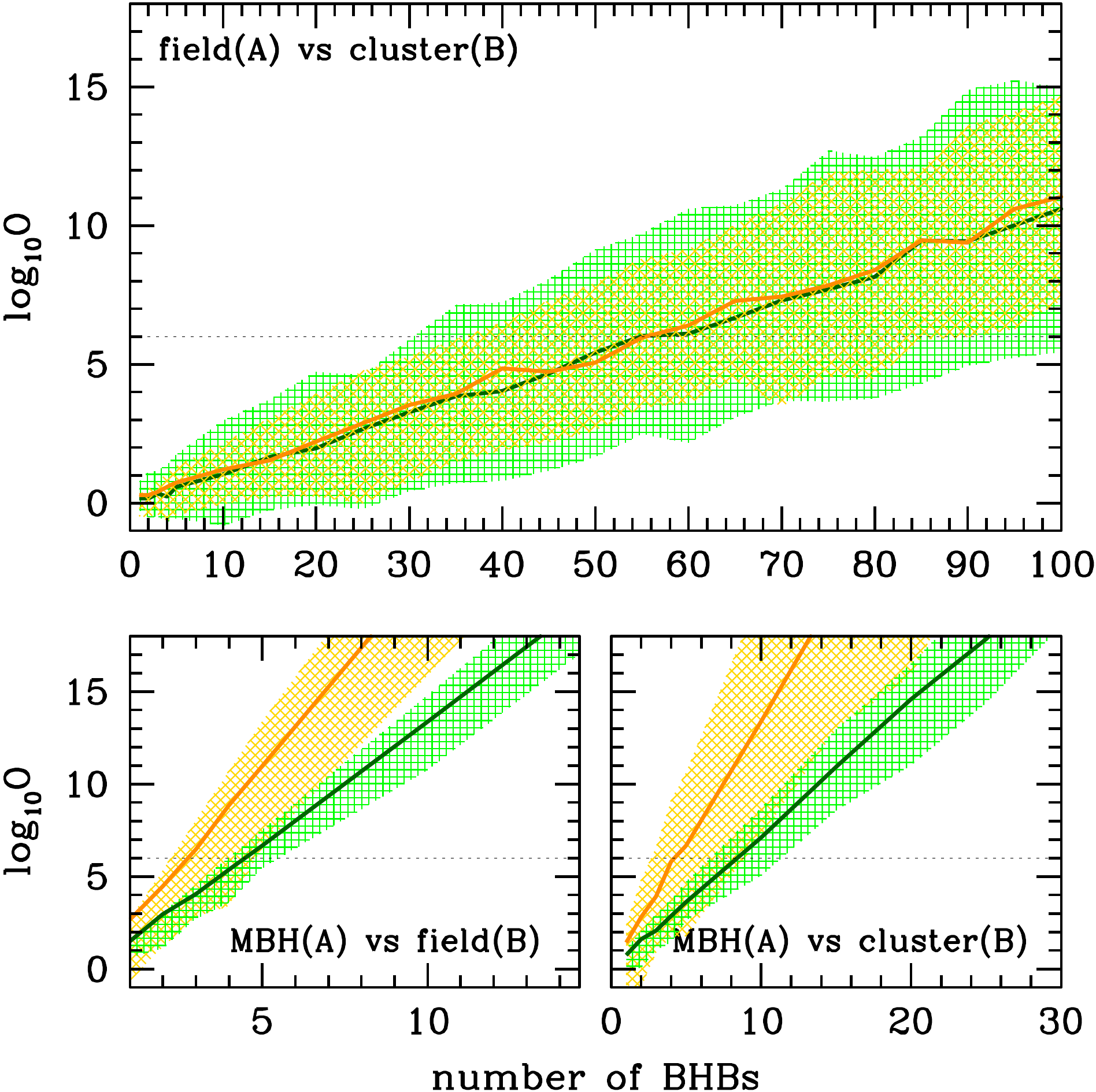}
\caption{Median and 90\% confidence interval of the odds ratio as a
  function of $N_{\rm obs}$ for baseline N2A2-5y. For each value of
  $N_{\rm obs}$, we consider 100 Monte Carlo realizations. In each
  comparison panel, orange curves and shaded areas represent $O_{AB}$
  when $A$ is the true model, whereas green curves represent $O_{BA}$
  when $B$ is the true model.}
\label{fig:oddsN2A2}
\end{center}
\end{figure} 

Since all detector baselines yield similar results, we take a closer
look at the N2A2-5y case, a plausible ``minimum'' eLISA baseline
target. In Figure~\ref{fig:oddsN2A2} we show odds ratios for this
specific configuration, including the 90\% confidence interval
computed from the 100 catalogues constructed for each $N_{\rm obs}$.
The log$(O)=6$ threshold is always achieved with less than 10 observed
BHBs when the {\it MBH} model is involved in the comparison, whereas
up to 100 BHBs may be needed to discriminate between the {\it field}
and {\it cluster} models, depending on the specific ensemble of
observed BHBs.

\begin{figure}
\begin{center}
\includegraphics[width=7.5cm]{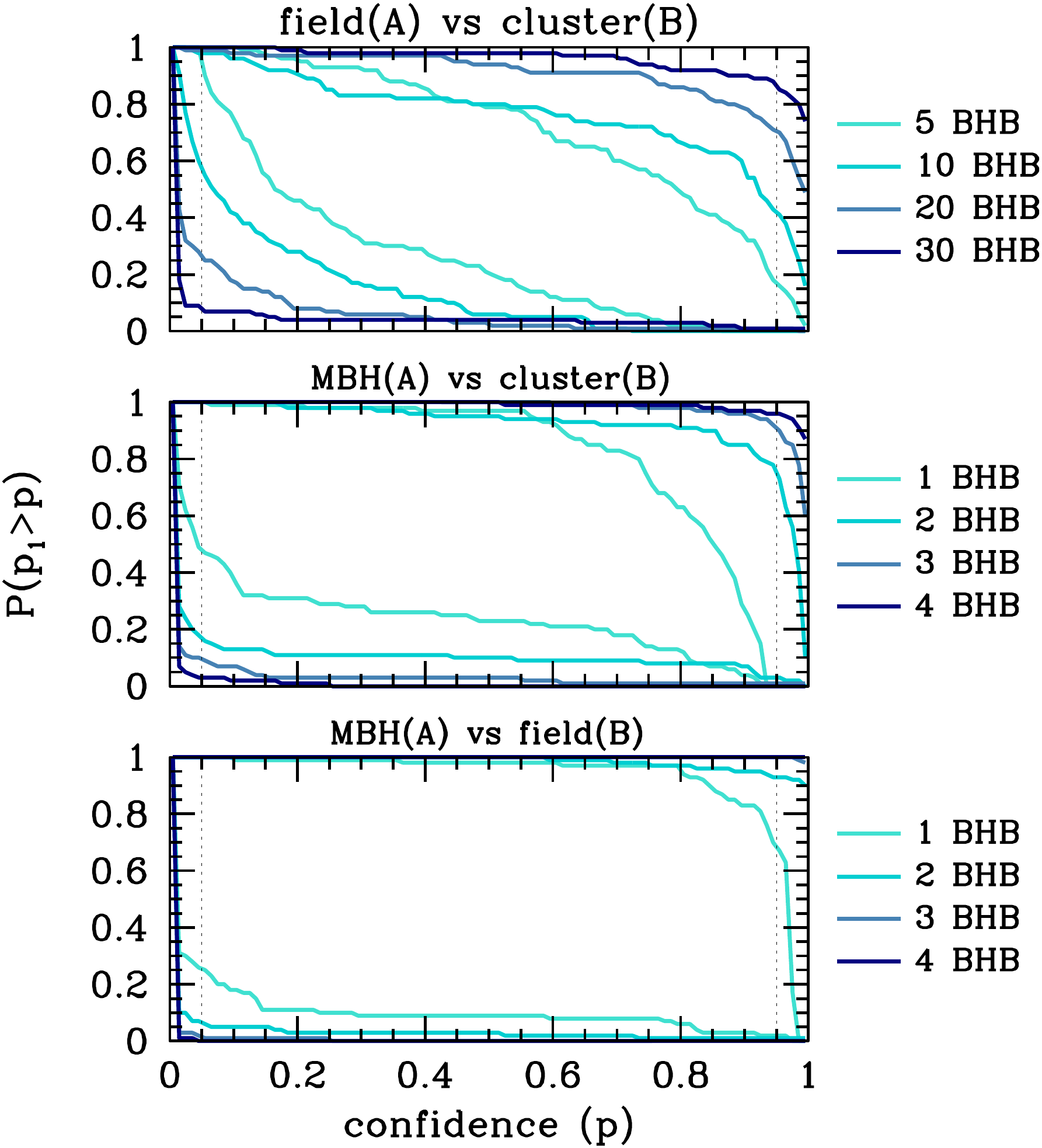}
\caption{CDF of the confidence in a given model over 100 Monte Carlo
  realizations of the observed BHB sample for different BHB
  observations $N_{\rm obs}$ (as labeled in each panel) assuming the
  N2A2-5y eLISA baseline. The top curve in each pair marks the CDF of
  confidence in model $A$ when model $A$ is true, whereas the bottom
  curve marks the CDF of confidence in model $A$ when model $B$ is
  true.}
\label{fig:pdist}
\end{center}
\end{figure} 

When comparing two models $A$ and $B$ for a given eLISA baseline, we
can use the 100 catalogues at fixed $N_{\rm obs}$ to construct
cumulative distributions functions (CDFs) of $p_A$ and $p_B$
\citep{2011PhRvD..83d4036S}. Suppose that, in a comparison between
models $A$ and $B$, $A$ is the right model, and we draw 100
realizations of $N_{\rm obs}$ observations from model A. we can
compute the associated CDF of $p_A$ and plot it against the confidence
($0<p<1$). The result for different values of $N_{\rm obs}$ is shown
in the upper curves of each panel of Figure~\ref{fig:pdist}. We can
also draw 100 realizations of $N_{\rm obs}$ observations from $B$ and
compute the CDF of $p_A$ when $A$ is not true. The result are the
lower curves in each panel of Figure~\ref{fig:pdist}.

Set, for example, $p=0.95$ (approx $2\sigma$). The value of the upper
curve at $p$ is the fraction of realizations for which we have more
than $2\sigma$ confidence that model A is correct when it is, in
fact, true. The value of the lower curve at $1-p$ is the fraction of
realization for which we cannot rule out model $A$ at $2\sigma$
confidence when it is the wrong model (i.e., observations are
generated by model $B$). Figure~\ref{fig:pdist} presents a similar
analysis for all pairs of models assuming the N2A2-5y baseline. About
30 BHB observations are required for a $2\sigma$--confidence
identification of model {\it field} against model {\it cluster} in
about 90\% of the realizations. The same level of confidence requires
only 4 and 2 BHBs when model {\it MBH} is compared to models {\it
  cluster} and {\it field}, respectively.

\section{Discussion and outlook}
\begin{table}
\begin{center}
\begin{tabular}{|c|c|cc|cc|}
\hline
&&\multicolumn{2}{|c|}{3$\sigma$}&\multicolumn{2}{|c|}{5$\sigma$}\\
eLISA base & $N_{\rm obs}$ & $N_{50}$ & $N_{90}$ & $N_{50}$ & $N_{90}$\\
\hline
N2A2-2y & 11-78    & 35 & $>$100 & 95 & $>$100\\
N2A5-2y & 85-595   & 34 & 95     & 80 & $>$100\\
N2A2-5y & 45-310   & 25 & 60     & 61 & 100\\
N2A5-5y & 330-2350 & 25 & 62     & 60 & 100\\

\hline
\hline 
\end{tabular}
\caption{Expected number of sources (column 2) for each eLISA baseline
  (column 1), compared with the number of observations needed to
  distinguish between models {\it field} and {\it cluster} at a given
  confidence threshold in 50\% ($N_{50}$) and 90\% ($N_{90}$) of the
  cases (columns 3-6).}
\label{tab1}
\end{center}
\end{table}

For the log-flat distribution assumed here, the Advanced LIGO
observations imply a 90\% credible interval for the merger rate of
${\cal R}=[10,70]$~yr$^{-1}$Gpc$^{-3}$ \citep{2016arXiv160604856T}. The
resulting range in $N_{\rm obs}$ is reported in Table~\ref{tab1} for
the different baselines, and it should be compared to the number of
events needed to discriminate among different models at a desired
confidence threshold. Model {\it MBH} can be identified by all the
configurations with just a few BHB observations, therefore it is not
reported in the table. Discriminating between the {\it cluster} and
{\it field} scenarios requires tens of events, and only the baseline
N2A5-5y can guarantee a 5$\sigma$ confidence with 90\%
probability. Baselines N2A2-5y and N2A5-2y can distinguish among these
models at the 3$\sigma$ level, but this may not be possible should the
event rate lean toward the lower limit of the allowed range. The
N2A2-2y baseline performs relatively poorly, and it may not deliver
enough detections to pin down the formation mechanism.

These results highlight the importance of aiming for a five-year
mission with the longest possible armlength. However, we should bear
in mind some limitations of our proof-of-principle analysis. First of
all, we selected three representative models from the literature: this
does not fully capture all of the relevant physics affecting the
eccentricity distribution of BHBs. For example, several variations of
the ``fiducial'' model of \cite{Kowalska:2010qg} result in slightly
different eccentricity distributions. Our analysis can be applied
systematically to any such variation, assessing to what extent the
underlying physics can be constrained. Secondly, we assumed the
eccentricity distribution to be independent of masses and
redshifts. In practice, different formation channels will result in
different mass-eccentricity (and possibly redshift-eccentricity, or
spin-eccentricity) correlations, that can be exploited in a
multi-dimensional analysis to enhance the discriminating power of the
observations. Finally, it is very likely that several different
formation channels operate at the same time in the Universe. In the
context of massive BHB observations, \cite{2011PhRvD..83d4036S}
studied whether eLISA could identify a superposition of distinct
formation channels from the statistical properties of the observed
population. A similar analysis in the present context is an
interesting topic for future work.

\section*{Acknowledgements}
A.S. thanks T. Dent and A. Vecchio for useful discussions.
E.B., A.K. and A.N. are supported by NSF CAREER Grant No.
  PHY-1055103.  E.B. and A.K. are supported by FCT contract
  IF/00797/2014/CP1214/CT0012 under the IF2014 Programme. This work was 
supported by the H2020-MSCA-RISE- 2015 Grant No. StronGrHEP-690904.  A.S. is
  supported by a University Research Fellowship of the Royal Society.

\bibliographystyle{mnras}

\bsp
\label{lastpage}
\end{document}